\documentstyle[twocolumn,prd,aps]{revtex}
\input{epsf}
\begin{document}
\title{
Renormalizability and model independent description of $Z^\prime$
signals at low energies}
\author{
 A. V. Gulov \thanks{Email address: gulov@ff.dsu.dp.ua} and
 V. V. Skalozub \thanks{Email address: skalozub@ff.dsu.dp.ua}}
\address{
 Dniepropetrovsk State University, Dniepropetrovsk, 49050 Ukraine}
\date{\today}
\maketitle
\begin{abstract}
Model independent search for signals of heavy $Z^\prime$ gauge
bosons in low-energy four-fermion processes is analyzed. It is
shown that the renormalizability of the underlying theory
containing $Z^\prime$, formulated as a scattering in the field of
heavy virtual states, can be implemented in specific relations
between different processes. Considering the two-Higgs-doublet
model as the low-energy basis theory, the two types of $Z^\prime$
interactions with light particles are found to be compatible with
the renormalizability. They are called the Abelian and the
``chiral'' couplings. Observables giving possibility to uniquely
detect $Z^\prime$ in both cases are introduced.
\end{abstract}

\section{Introduction}\label{sec:introduction}

The existence of the heavy $Z^\prime$ gauge boson is predicted by
a number of grand unified theories (GUT's) and superstring
theories \cite{leike}. The mass of this particle is expected to be
of order $m_{Z^\prime}\geq 500$ GeV, and therefore it cannot be
produced at present day accelerators. Various strategies of
searching for signals of $Z^\prime$ as a virtual heavy state were
developed and different observables convenient for its
experimental detection have been introduced (see the survey
\cite{cvetic} and references therein). The model-dependent and
model-independent $Z^\prime$ searches at $e^+e^-$ colliders are
discussed (see, for instance, the report \cite{riemann}). A
popular model assumes that at low energies the $Z^\prime$
interactions with ordinary particles of the Standard Model (SM)
can be described by the effective gauge group ${\rm
SU}(2)_L\times{\rm U}(1)_Y\times{\rm \tilde{U}}(1)$. An
alternative choice is the gauge group ${\rm SU}(2)_L\times{\rm
SU}(2)_R\times{\rm U}(1)_{B-L}$ \cite{cvetic,sirlin}. These models
are considered as the remnants of underlying theories which are
not specified. The low-energy effective Lagrangians (EL) take into
consideration the most general property of renormalizable
theories, ensured by the decoupling theorem
\cite{decoupling,bando} -- the dominance of renormalizable
interactions at low energies. The interactions of
non-renormalizable types, being generated at high energies due to
radiation corrections, are suppressed by the inverse heavy mass
$1/m_{Z^\prime}$. Therefore, it is possible not to consider them
in leading order at lower energies. Another popular description is
the introduction of the EL, considered as the sum of all effective
operators with dimensions $n\ge 4$, constructed from the fields of
light particles. The coefficients at these operators are treated
as independent unknown numbers to be determined in experiments.
For more details see Ref. \cite{wudka}. In general, the number of
possible $Z^\prime$ couplings is large. So, it is difficult to
introduce observables allowing a unique detection of $Z^\prime$
signals. In this regard, it is desirable either to decrease the
number of the independent $Z^\prime$ parameters on some reasonable
grounds and to introduce observables most sensitive to the
$Z^\prime$ virtual states. In any case, the main idea is to find
correlations between the $Z^\prime$ couplings at low energies.

A straightforward way to find the correlations is to specify the
underlying theory describing interactions at energies
$\sim\Lambda_{\rm GUT}$ and to consider running of the couplings
from high to low energies $\sim m_W$ by using the renormalization
group (RG) equations. In this approach, each underlying theory
leads to the unique values of the parameters and, hence, the
corresponding correlations are model dependent ones. Another way
is to specify a basis low-energy theory (for instance, the SM can
be chosen) and to determine the relations between the $Z^\prime$
parameters, following from some model independent arguments. These
correlations are to be model independent. Naturally, they remain
dependent on the chosen basis low-energy theory.

In Refs. \cite{Zpr,obs} the method for derivation model
independent correlations between the parameters of physics beyond
the SM has been developed, and new observables convenient in
searching for the $Z^\prime$ boson in four-fermion processes were
introduced. This approach is based on principles of the RG and the
decoupling theorem \cite{decoupling}. As it was argued, any
virtual heavy particle can be treated as an ``external field''
scattering the SM particles. The vertex describing interaction
with the field contains a numeric factor, which is considered as
an arbitrary parameter. Actually, it is generated by the
decoupling and therefore depends on the underlying model. Due to
renormalizability, the scattering amplitude in the ``external
field'' satisfies some simple relation (named RG relation), which
includes the $\beta$ and $\gamma$ functions entering the RG
equation. These functions have to be calculated with the light
particles only, and the vertex factor. Hence, relations between
different vertex factors follow. Then, they can be implemented in
a number of model independent observables corresponding to the
specific heavy virtual state, in particular, to the $Z^\prime$
gauge boson \cite{obs}.

In Ref. \cite{Zpr} as the low-energy basis model the minimal SM
(with one scalar doublet) has been chosen. However, at present
there is a few information about the scalar fields. In this
regard, the theory with two scalar doublets is intensively studied
\cite{hhguide,santos}. The two-Higgs-doublet model (THDM) is also
known as the low-energy limit of some ${\rm E}_6$ based GUT's,
which predict the $Z^\prime$ gauge boson. In the present paper,
the results of Ref. \cite{Zpr} are generalized to the THDM case.
We analyse in detail both the Abelian and the so called ``chiral''
types of the $Z^\prime$ couplings to light particles. As the
latter type is concerned, it was derived as follows. We first
assumed the most general parametrization of $Z^\prime$
interactions with the SM fields and then derived the generator
structures, compatible with the renormalizability. As it will be
shown in what follows, there is an important difference between
these two types of interactions.

Thus, in order to derive the model independent constraints we
choose the THDM as the low-energy basis theory (notice, the
minimal SM is a particular case of the THDM). Then, we introduce a
general parametrization of linear in $Z^\prime$ couplings, which
is independent of the specific underlying theory. As a result, the
derived RG correlations are model independent ones. They hold for
the THDM as well as for the minimal SM. Moreover, the existence of
other heavy particles with masses $m_i\ge m_{Z^\prime}$ does not
affect these correlations.

As it will be shown, there are two completely different sets of
the $Z^\prime$ couplings to the SM fields compatible with
renormalizability. The first one describes the Abelian $Z^\prime$,
which respects the additional ${\rm \tilde{U}}(1)$ symmetry of the
low energy EL. In this case the $Z^\prime$ couplings to the
axial-vector fermion currents have a universal absolute value. The
second set corresponds to the chiral $Z^\prime$, which interacts
with the SM doublets, only. One has to distinguish these neutral
$Z^\prime$ gauge bosons because they are described by different
operators.

The content is as follows. In Sec. \ref{sec:model} the general
parametrization of interactions involving the $Z^\prime$ and the
SM fields is introduced. The RG correlations between the
$Z^\prime$ couplings are derived in Sec. \ref{sec:RGrelations}. In
Sec. \ref{sec:e6} they are compared with the specific values of
the $Z^\prime$ couplings in the GUT's based on the ${\rm E}_6$
group. In Sec. \ref{sec:observ} the observables convenient in
detection of the $Z^\prime$ signals are proposed. The results of
our investigation are discussed in Sec. \ref{sec:discussion}.

\section{Parametrization of the $Z^\prime$ couplings}\label{sec:model}

In the present paper we analyze the four-fermion scattering
amplitudes of order $\sim m^{-2}_{Z^\prime}$ generated by the
virtual $Z^\prime$ states. Vertices of interactions with more than
one $Z^\prime$ field contribute to the amplitudes involving
several virtual $Z^\prime$ states. The latter processes have order
$m^{-4}_{Z^\prime}$ and higher. Therefore, in what follows we
consider the linear in $Z^\prime$ vertices, only.

To introduce a general parametrization of the vertices involving
the SM fields and being linear in the $Z^\prime$ field, let us
impose a number of natural conditions. First of all, the
renormalizable type interactions are dominant at low energies
$\sim m_W$. The non-renormalizable interactions generated at high
energies due to radiation corrections are suppressed by the
inverse heavy mass $1/m_{Z^\prime}$ (or by other heavier scales
$1/\Lambda_i\ll 1/m_{Z^\prime}$) and therefore at low energies can
be neglected in leading order. We assume that the $Z^\prime$ is
the only neutral vector boson with the mass $\sim m_{Z^\prime}$,
and the $Z^\prime$ gauge field enters the theory through covariant
derivatives with a corresponding charge. We also assume that the
${\rm SU}(2)_L\times{\rm U}(1)_Y$ gauge group of the SM is a
subgroup of the GUT group. In this case, a product of generators
associated with the SM subgroup is a linear combination of these
generators. As a consequence, the all structure constants
connecting two SM gauge bosons with $Z^\prime$ have to be zero.
Hence, the interactions of gauge fields of the types $Z^\prime W^+
W^-$, $Z^\prime Z Z$, and other are absent at tree level.

Let $\phi_i$ ($i=1,2$) be two complex scalar doublets:
\begin{equation}
\phi_i=\left\{a^+_i, \frac{v_i +b_i +i c_i}{\sqrt{2}}\right\},
\end{equation}
where $v_i$ marks corresponding vacuum expectation values, $a^+_i$
are complex fields, and $b_i$, $c_i$ are real fields. By
diagonalizing the quadratic terms of the scalar potential
$V(\phi_1,\phi_2)$ one obtains the mass eigenstates: two neutral
$CP$-even scalar particles, $H$ and $h$, the neutral $CP$-odd
scalar particle, $A_0$, the Goldstone boson partner of the $Z$
boson, $\chi_3$, the charged Higgs field, $H^\pm$, and the
Goldstone field associated with the $W^\pm$ boson, $\chi^\pm$:
\begin{eqnarray}\label{scalars:rule}
 a^+_1 =& \chi^+\cos\beta -H^+\sin\beta,\quad&
 a^+_2 = H^+\cos\beta +\chi^+\sin\beta, \nonumber\\
 c_1 =& \chi_3\cos\beta -A_0\sin\beta,\quad&
 c_2 = A_0 \cos\beta +\chi_3 \sin\beta, \nonumber\\
 b_1 =& H\cos\alpha -h\sin\alpha,\quad&
 b_2 = h\cos\alpha +H\sin\alpha,
\end{eqnarray}
where
\begin{equation}
\tan\beta =\frac{v_2}{v_1},
\end{equation}
and the angle $\alpha$ is determined by the explicit form of the
potential $V(\phi_1,\phi_2)$. For instance, the $CP$-conserving
potential, which has only $CP$-invariant minima, can be used
\cite{hhguide,santos}:
\begin{eqnarray}\label{L:potential}
V&=& \sum\limits_{i=1}^2\left[
  -\mu^2_i \phi^\dagger_i\phi_i
  +\lambda_i(\phi^\dagger_i \phi_i)^2\right]
 +\lambda_3 (\mbox{Re}[\phi^\dagger_1 \phi_2])^2
 \nonumber\\&&
 +\lambda_4 (\mbox{Im}[\phi^\dagger_1 \phi_2])^2
 +\lambda_5 (\phi^\dagger_1 \phi_1) (\phi^\dagger_2 \phi_2).
\end{eqnarray}
It is consistent with the absence of the tree-level
flavor-changing neutral currents (FCNC's) in the fermion sector.
The corresponding value of $\alpha$ is \cite{santos}
\begin{equation}
\tan 2\alpha = -\frac{v_1 v_2 \left(\lambda_3+\lambda_5\right)}
{\lambda_2 v^2_2-\lambda_1 v^2_1}.
\end{equation}

At low energies, when all heavy states are decoupled, the
$Z^\prime$ interactions with the scalar doublets can be
parametrized in a model independent way as follows \cite{cvetic}:
\begin{eqnarray}\label{L:scalar}
 {\cal L}_\phi
 &=&\sum\limits_{i=1}^2
 \left|\left(\partial_\mu
 -\frac{ig}{2}\sigma_a W^a_\mu
 -\frac{i{g^\prime}}{2}Y_{\phi_i} B_\mu
\right.\right.\nonumber\\&&\left.\left.
 - \frac{i\tilde{g}}{2}\tilde{Y}_{\phi_i} \tilde{B}_\mu
 \right)\phi_i\right|^2,
\end{eqnarray}
where $g$, $g^\prime$, $\tilde{g}$ are the charges associated with
the ${\rm SU}(2)_L$, ${\rm U}(1)_Y$, and the $Z^\prime$ gauge
groups, respectively, $\sigma_a$ are the Pauli matrices,
\begin{eqnarray}
\tilde{Y}_{\phi_i}&=& \left(\begin{array}{cc}
 \tilde{Y}_{\phi_i,1} & 0 \\ 0 & \tilde{Y}_{\phi_i,2}
 \end{array}\right)
\end{eqnarray}
is the generator corresponding to the gauge group of the
$Z^\prime$ boson, and $Y_{\phi_i}$ is the ${\rm U}(1)_Y$
hypercharge. The condition $Y_{\phi_i} =1$ guarantees that the
vacuum is invariant with respect to the gauge group of photon.

The vector bosons, $A$, $Z$, and $Z^\prime$, are related with the
symmetry eigenstates as follows:
\begin{eqnarray}\label{vectors:rule}
 B&\to& A\cos{\theta_W}-(Z\cos{\theta_0}
  -Z^\prime\sin{\theta_0})\sin{\theta_W},
  \nonumber\\
 W_3&\to& A\sin{\theta_W} +(Z\cos{\theta_0}
  -Z^\prime\sin{\theta_0})\cos{\theta_W},
  \nonumber\\
 \tilde{B}&\to& Z\sin{\theta_0}
 +Z^\prime\cos{\theta_0},
\end{eqnarray}
where $\tan\theta_W=g^\prime/g$ is the adopted in the SM value of
the Weinberg angle, and
\begin{eqnarray}\label{ZZpmixing}
 \tan{\theta_0}&=&
 \frac{\tilde{g} m^2_W
 \left(\tilde{Y}_{\phi_1,2}\cos^2\beta
 +\tilde{Y}_{\phi_2,2}\sin^2\beta\right)}
 {g\cos{\theta_W}
 \left(m^2_{Z^\prime}-m^2_W/\cos^2{\theta_W}\right)}.
\end{eqnarray}
As is seen, the mixing angle $\theta_0$ is of order $\sim
m^2_W/m^2_{Z^\prime}$. That results in the corrections of order
$\sim m^2_W/m^2_{Z^\prime}$ to the interactions between the SM
particles. To avoid the tree-level mixing of the $Z$ boson and the
physical scalar field $A_0$ one has to impose the condition
$\tilde{Y}_{\phi_1,2}=
\tilde{Y}_{\phi_2,2}\equiv\tilde{Y}_{\phi,2}$.

Now, let us parametrize the fermion-vector interactions
introducing the effective low-energy Lagrangian
\cite{cvetic,sirlin,caso}:
\begin{eqnarray}\label{L:fermion}
 {\cal L}_f
&=&i\sum\limits_{f_L}\bar{f}_L{\gamma^\mu}
 \Big(\partial_\mu
 -\frac{ig}{2}{\sigma_a}W^a_\mu
 -\frac{i g^\prime}{2}{B_\mu}Y_{f_L}
 \nonumber\\&&
 -\frac{i\tilde{g}}{2}\tilde{B}_\mu\tilde{Y}_{f_L}\Big)f_L
 \nonumber\\
 &&+i\sum\limits_{f_R}\bar{f}_R{\gamma^\mu}
 \Big(\partial_\mu -i g^\prime B_\mu Q_f
 -\frac{i\tilde{g}}{2}\tilde{B}_\mu\tilde{Y}_{R,f}\Big)f_R,
\end{eqnarray}
where the renormalizable type interactions are admitted and the
summation over the all SM left-handed fermion doublets, $f_L
=\{(f_u)_L, (f_d)_L\}$, and the right-handed singlets, $f_R =
(f_u)_R, (f_d)_R$, is understood. $Q_f$ denotes the charge of $f$
in the positron charge units,
\begin{eqnarray}
 \tilde{Y}_{f_L}&=&
 \left(\begin{array}{cc}
 \tilde{Y}_{L,f_u} & 0 \\ 0 & \tilde{Y}_{L,f_d} \end{array}\right),
\end{eqnarray}
and $Y_{f_L}$ equals to $-1$ for leptons and $1/3$ for quarks.

Renormalizable interactions of fermions and scalars are described
by the Yukawa Lagrangian. To avoid the existence of the tree-level
FCNC's one has to ensure that at the diagonalization of the
fermion mass matrix the diagonalization of the scalar-fermion
couplings is automatically fulfilled. In this case the Yukawa
Lagrangian, which respects the ${\rm SU}(2)_L\times{\rm U}(1)_Y$
gauge group, can be written in the form:
\begin{eqnarray}\label{L:Yuk}
{\cal L}_{\rm Yuk} &=&
 -\sqrt{2}\sum\limits_{f_L}\sum\limits_{i=1}^{2}\left\{
 G_{f_d,i}\left[
  \bar{f}_L\phi_i(f_d)_R +(\bar{f}_d)_R \phi^\dagger_i f_L
 \right]
 \right.\nonumber\\&&\left.
+G_{f_u,i}\left[
  \bar{f}_L\phi^c_i(f_u)_R +(\bar{f}_u)_R \phi^{c\dagger}_i f_L
 \right] \right\},
\end{eqnarray}
where $\phi^c_i=i\sigma_2\phi^\ast_i$ is the charge conjugated
scalar doublet, and the Cabibbo-Kobayashi-Maskawa mixing is
neglected. Then, the fermion masses are
\begin{equation}
m_f =\frac{2m_W}{g}\left(G_{f,1}\cos\beta
+G_{f,2}\sin\beta\right).
\end{equation}

As was shown by Glashow and Weinberg \cite{FCNC}, the tree-level
FCNC's mediated by Higgs bosons are absent in case when all
fermions of a given electric charge couple to no more than one
Higgs doublet. This restriction leads to four different models, as
discussed in Ref. \cite{santos}. In what follows, we will use the
most general parametrization (\ref{L:Yuk}) including the models
mentioned as well as other possible variations of the Yukawa
sector without the tree-level FCNC's.

By using Eqs. (\ref{L:scalar}), (\ref{L:fermion}), and
(\ref{L:Yuk}) it is easy to derive the Feynman rules which are
collected in Appendix \ref{sec:vertices}.

\section{RG relations}\label{sec:RGrelations}

In this section we consider the correlations between the
parameters $\tilde{Y}_{L,f}$, $\tilde{Y}_{R,f}$,
$\tilde{Y}_{\phi_i,1}$, and $\tilde{Y}_{\phi_i,2}$ appearing due
to the renormalizability of an underlying theory.

As is known, $S$-matrix elements are to be invariant with respect
to the RG transformations, which express the independence of the
location of a normalization point $\kappa$ in the momentum space.
In a theory with different mass scales the decoupling of heavy
loop contributions at the thresholds of heavy masses, $\Lambda$,
results in the important property of low energy amplitudes: the
running of all functions is regulated by the loops of light
particles. Therefore, the $\beta$ and $\gamma$ functions at low
energies are determined by the SM particles, only. This fact is
the consequence of the decoupling theorem \cite{decoupling}.

Actually, the decoupling results in the redefinition of the
parameters of the theory at the scale $\Lambda$ and removing the
all heavy particle loop contributions proportional to $\ln\kappa$
from the RG equation \cite{bando,2sc}:
\begin{eqnarray}\label{decoupling}
\lambda_a&=&\hat{\lambda}_a
+a_{\lambda_a}\ln\frac{\hat{\Lambda}^2}{\kappa^2}
+b_{\lambda_a}\ln^2\frac{\hat{\Lambda}^2}{\kappa^2} +...,
\nonumber\\ X&=&\hat{X}\left( 1
+a_X\ln\frac{\hat{\Lambda}^2}{\kappa^2}
+b_X\ln^2\frac{\hat{\Lambda}^2}{\kappa^2} +...\right),
\end{eqnarray}
where we use the notation $\lambda_a$ to refer to the charges, and
$X$ represents the all fields and masses. Hats over quantities
mark the parameters of the underlying theory. They include the
loops of both the SM and the heavy particles, whereas the
quantities without hats are calculated assuming that no heavy
particles are excited inside loops. The matching between the both
sets of parameters ($\lambda_a$, $X$ and $\hat{\lambda}_a$,
$\hat{X}$) is chosen to be done at the normalization point
$\kappa\sim\Lambda$,
\begin{equation}
\lambda_a\mid_{\kappa=\Lambda}
 =\hat{\lambda}_{a}\mid_{\kappa=\Lambda},
\quad X\mid_{\kappa=\Lambda}=\hat{X}\mid_{\kappa=\Lambda}.
\end{equation}
Since the sets of parameters $\lambda_a$, $X$ and
$\hat{\lambda}_a$, $\hat{X}$ differ at one-loop level, it is
possible to substitute one set by another.

As is shown in Ref. \cite{Zpr}, the redefinition of fields and
charges (\ref{decoupling}) allows one to eliminate the one-loop
mixing between heavy and light virtual states. Therefore, virtual
states of heavy particles can be treated as the ``external
fields'' scattering SM particles. The renormalizability of the
underlying theory leads to some relations for vertices describing
this scattering, called the RG relations.

Let us consider the four-fermion amplitudes caused by the
$Z^\prime$ boson exchange. In the lower order in ratio
$m^2_W/m^2_{Z^\prime}$ the process $\bar{f}_1
f_1\to{Z^\prime}^\ast\to\bar{f}_2 f_2$ can be presented as
scattering of the initial, $f_1$, and the final, $f_2$, fermions
in the ``external field'' $1/m_{Z^\prime}$ with the corresponding
vertex factors $\Gamma_{f_1 Z^\prime}$, $\Gamma_{f_2 Z^\prime}$.
The quantity $\Gamma_{f Z^\prime}$ contains no contributions of
heavy particle loops. Thus, it can be computed as a linear
combination of the parameters $\tilde{Y}_{L,f}$,
$\tilde{Y}_{R,f}$, $\tilde{Y}_{\phi_i,1}$, and
$\tilde{Y}_{\phi_i,2}$.

The RG invariance of the vertex leads to equation
\begin{equation}\label{RGE}
{\cal D}\left(\bar{f}\Gamma_{f
Z^\prime}f\frac{1}{m_{Z^\prime}}\right)=0,
\end{equation}
where the effective low-energy RG operator \cite{bando} is defined
as follows:
\begin{eqnarray}
{\cal D}&\equiv& \frac{\partial}{\partial\ln\kappa}
+\sum\limits_{a}{\beta_a}\frac{\partial}{\partial{\lambda_a}}
-\sum\limits_{X}{\gamma_X}\frac{\partial}{\partial\ln{X}},
\nonumber\\ \beta_a&=&\frac{d\lambda_a}{d\ln\kappa},\quad
\gamma_X=-\frac{d\ln{X}}{d\ln\kappa},
\end{eqnarray}
and the coefficient functions $\beta_a$ and $\gamma_X$ are
computed taking into account the loops of light particles.

Relation (\ref{RGE}) ensures that, as a consequence of
renormalizability, the mathematical structure of the leading
logarithmic terms of the vertices, calculated in one- and
higher-loop approximations, coincides with that of the tree-level
structures. The standard usage of Eq. (\ref{RGE}) is to improve
scattering amplitudes calculated in a fixed order of perturbation
theory. In contrast, in what follows we will apply Eq. (\ref{RGE})
to obtain an algebraic relation between the parameters
$\tilde{Y}_{L,f}$, $\tilde{Y}_{R,f}$, $\tilde{Y}_{\phi_i,1}$,
$\tilde{Y}_{\phi_i,2}$, which are to be considered as arbitrary
numbers, since the underlying theory is not specified. Let us
explain the idea in more detail. In case when the underlying
theory is specified ($\tilde{Y}_{L,f}$, $\tilde{Y}_{R,f}$,
$\tilde{Y}_{\phi_i,1}$, $\tilde{Y}_{\phi_i,2}$ have to be computed
as discussed before), and the $\beta$ and $\gamma$ functions as
well as the $S$-matrix elements are calculated in a fixed order of
perturbation theory, Eq. (\ref{RGE}) is just the identity. If the
underlying theory is not specified, whereas the $\beta$, $\gamma$
functions and $S$-matrix elements are computed in a fixed order of
perturbation theory, equality (\ref{RGE}) may serve to correlate
the unknown parameters $\tilde{Y}$. In case of the four-fermion
processes mediated by the gauge $Z^\prime$ boson, the number of
independent $\beta$ functions is less than the number of RG
equations. Therefore, the non-trivial system of equations
correlating the originally independent parameters occurs.

The one-loop RG relation for the fermion-$Z^\prime$ vertex is
\cite{Zpr}
\begin{equation}\label{RGrelation}
\bar{f}\frac{\partial\Gamma^{(1)}_{f
Z^\prime}}{\partial\ln\kappa}f \frac{1}{m_{Z^\prime}} +{\cal
D}^{(1)}\left(\bar{f}\Gamma^{(0)}_{f Z^\prime}f
\frac{1}{m_{Z^\prime}}\right) =0,
\end{equation}
where $\Gamma^{(0)}_{f Z^\prime}$ and $\Gamma^{(1)}_{f Z^\prime}$
denote the tree-level and the one-loop level contributions to the
fermion-$Z^\prime$ vertex, and ${\cal D}^{(1)}$ is the one-loop
level part of the RG operator,
\begin{equation}
{\cal D}^{(1)}\equiv
\sum\limits_a{\beta}^{(1)}_a\frac{\partial}{\partial\lambda_a}-
\sum\limits_{X}{\gamma}^{(1)}_{X}\frac{\partial}{\partial\ln{X}}.
\end{equation}

\begin{figure}
\begin{center}
 \epsfxsize=0.4\textwidth
 \epsfbox[10 4 577 705]{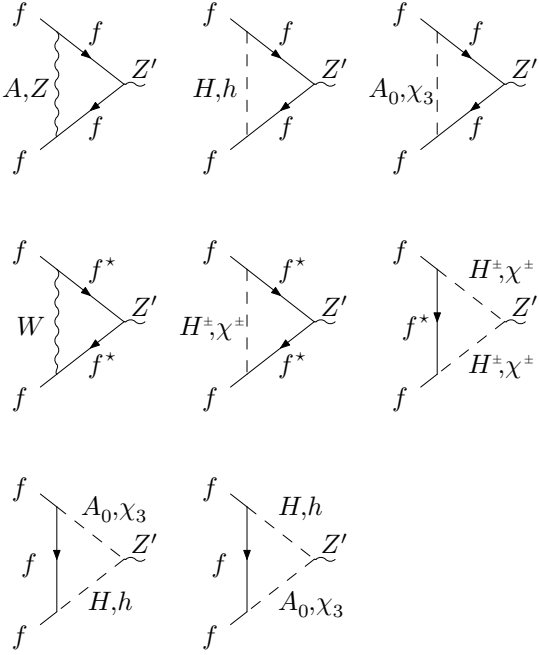}
 \caption{One-loop contributions to the divergent part
  of $\Gamma_{f Z^\prime}$.}
 \label{fig:1}
 \end{center}
\end{figure}

As it follows from Eq. (\ref{RGrelation}), only the divergent
parts of the one-loop vertices $\Gamma^{(1)}_{f Z^\prime}$ are to
be calculated. The corresponding diagrams are shown in Fig.
\ref{fig:1}. The following expressions for the right-handed and
the left-handed fermions, respectively, have been obtained,
\begin{eqnarray}
\frac{\partial\Gamma^\mu_{f_R Z^\prime}}{\partial\ln\kappa}
 &=&\frac{\gamma^\mu}{8 \pi^2}
 \left\{g^2 Q^2_f \tilde{Y}_{R,f} \tan^2{\theta_W}
  +\frac{4}{3}g^2_{s,f}\tilde{Y}_{R,f}\right.
 \nonumber\\&&
  +G^2_{f,1}\left[
   2T^3_f\left(\tilde{Y}_{\phi,2} +\tilde{Y}_{\phi_1,1}\right)
   +\tilde{Y}_{L,f} +\tilde{Y}_{L,f^\star}
  \right]
 \nonumber\\&&
  +G^2_{f,2}\left[
   2T^3_f\left(\tilde{Y}_{\phi,2} +\tilde{Y}_{\phi_2,1}\right)
   +\tilde{Y}_{L,f} +\tilde{Y}_{L,f^\star}
  \right]
 \nonumber\\&& \left.
 +O\left(\frac{m^2_W}{m^2_{Z^\prime}}\right)
 \right\},
 \nonumber\\
\frac{\partial\Gamma^\mu_{f_L Z^\prime}}{\partial\ln\kappa}
 &=&\frac{\gamma^\mu}{8 \pi^2}
  \left\{\frac{g^2}{2}\tilde{Y}_{L,f^\star}
  +\frac{4}{3}g^2_{s,f}\tilde{Y}_{L,f}\right.
 \nonumber\\&&
  +g^2\tilde{Y}_{L,f}\left[
   \frac{1}{4\cos^2{\theta_W}}
   +\left(Q^2_f -\left|Q_f\right|\right)\tan^2{\theta_W}
  \right]
 \nonumber\\&&
  +\left(G^2_{f,1}+G^2_{f,2}\right)
   \left(\tilde{Y}_{R,f} -2T^3_f\tilde{Y}_{\phi,2}\right)
 \nonumber\\&&
  +G^2_{f^\star,1}\left(2T^3_f\tilde{Y}_{\phi_1,1}
   +\tilde{Y}_{R,f^\star}\right)
 \nonumber\\&&
  +G^2_{f^\star,2}\left(2T^3_f\tilde{Y}_{\phi_2,1}
   +\tilde{Y}_{R,f^\star}\right)
 \nonumber\\&& \left.
 +O\left(\frac{m^2_W}{m^2_{Z^\prime}}\right)\right\},
\end{eqnarray}
where $f$ and $f^\star$ are the partners of a ${\rm SU}(2)_L$
fermion doublet (namely, $l^\star=\nu_l$, $\nu^\star_l=l$,
$q^\star_u=q_d$, and $q^\star_d=q_u$), $T^3_f$ is the third
component of the weak isospin, and $g_{s,f}$ is the QCD charge for
quarks, and zero for leptons.

\begin{figure}
\begin{center}
 \epsfxsize=0.4\textwidth
 \epsfbox[10 18 581 470]{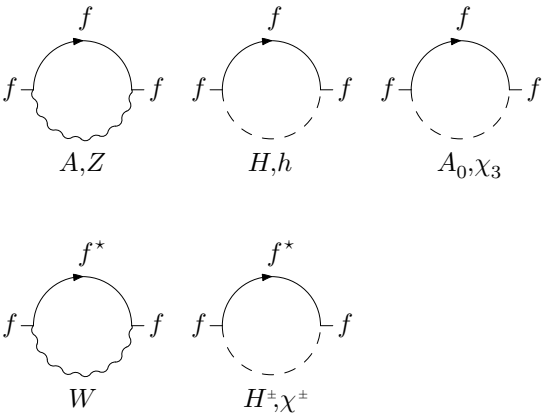}
 \caption{One-loop contributions to the fermion mass operator.}
 \label{fig:2}
 \end{center}
\end{figure}

The fermion anomalous dimensions can be calculated by using the
diagrams of Fig. \ref{fig:2}:
\begin{eqnarray}
\gamma_{f_R}&=&\frac{1}{16 \pi^2}\left[
 g^2 Q^2_f \tan^2{\theta_W} +\frac{4}{3}g^2_{s,f}
 +2\left(G^2_{f,1} +G^2_{f,2}\right)
\right.\nonumber\\&&\left.
 +O\left(\frac{m^2_W}{m^2_{Z^\prime}}\right)
 \right],
\nonumber\\ \gamma_{f_L}&=&\frac{1}{16 \pi^2}\left[
 g^2\left(Q^2_f -\left|Q_f\right|\right)\tan^2{\theta_W}
 +\frac{4}{3}g^2_{s,f} +\frac{g^2}{2}
\right.\nonumber\\&&
 +\frac{g^2}{4\cos^2{\theta_W}}
 +G^2_{f,1} +G^2_{f,2} +G^2_{f^\star,1} +G^2_{f^\star,2}
 \nonumber\\&&\left.
 +O\left(\frac{m^2_W}{m^2_{Z^\prime}}\right)
 \right].
\end{eqnarray}

RG relations (\ref{RGrelation}) considered in a lower order in
$m^2_W/m^2_{Z^\prime}$ lead to the equations for the parameters
$\tilde{Y}_{L,f}$, $\tilde{Y}_{R,f}$, $\tilde{Y}_{\phi_i,1}$, and
$\tilde{Y}_{\phi_i,2}$:
\begin{eqnarray}\label{eq}
 &&4\pi^2\tilde{Y}_{R,f}
 \left(\frac{\beta^{(1)}_{\tilde{g}}}{\tilde{g}^2}
  +\gamma^{(1)}_{m^2_{Z^\prime}}\right)=
 \nonumber\\&&\quad
 -G^2_{f,1}\left[
  2T^3_f\left(\tilde{Y}_{\phi,2} +\tilde{Y}_{\phi_1,1}\right)
  +\tilde{Y}_{L,f} +\tilde{Y}_{L,f^\star} -2\tilde{Y}_{R,f}
 \right]
 \nonumber\\&&\quad
 -G^2_{f,2}\left[
  2T^3_f\left(\tilde{Y}_{\phi,2} +\tilde{Y}_{\phi_2,1}\right)
  +\tilde{Y}_{L,f} +\tilde{Y}_{L,f^\star} -2\tilde{Y}_{R,f}
 \right],
\nonumber \\
 &&4\pi^2\tilde{Y}_{L,f}
 \left(\frac{\beta^{(1)}_{\tilde{g}}}{\tilde{g}^2}
   +\gamma^{(1)}_{m^2_{Z^\prime}}\right)=
 \frac{g^2}{2}
  \left(\tilde{Y}_{L,f} -\tilde{Y}_{L,{f^\star}} \right)
 \nonumber\\&&\quad
 +\left(G^2_{f,1} +G^2_{f,2}\right)\left(
  2T^3_f\tilde{Y}_{\phi,2} +\tilde{Y}_{L,f} -\tilde{Y}_{R,f}
 \right)
 \nonumber\\&&\quad
 -G^2_{f^\star,1}\left(2T^3_f\tilde{Y}_{\phi_1,1}
  -\tilde{Y}_{L,f} +\tilde{Y}_{R,f^\star}\right)
 \nonumber\\&&\quad
 -G^2_{f^\star,2}\left(2T^3_f\tilde{Y}_{\phi_2,1}
  -\tilde{Y}_{L,f} +\tilde{Y}_{R,f^\star}\right).
\end{eqnarray}

One has to derive two sets of relations, which ensure the
compatibility of Eqs. (\ref{eq}). The first one is
\begin{eqnarray}\label{non-Abelian}
&&\tilde{Y}_{\phi_2,1} =\tilde{Y}_{\phi_1,1}
 =-\tilde{Y}_{\phi,2}\equiv-\tilde{Y}_\phi,
 \nonumber\\&&
\tilde{Y}_{L,f} +\tilde{Y}_{L,f^\star}=0,\quad \tilde{Y}_{R,f}=0.
\end{eqnarray}
It describes the $Z^\prime$ boson analogous to the third component
of the ${\rm SU}(2)_L$ gauge field. The characteristic features of
these interactions are the zero traces of generators and the
absence of couplings to the right-handed singlets. In what
follows, we shall call this type of interaction the ``chiral''
$Z^\prime$. The second set,
\begin{eqnarray}\label{Abelian}
&&\tilde{Y}_{\phi_1,1} =\tilde{Y}_{\phi_2,1} =\tilde{Y}_{\phi,2}
\equiv\tilde{Y}_{\phi}, \nonumber\\&& \tilde{Y}_{L,f}
=\tilde{Y}_{L,f^\star},\quad \tilde{Y}_{R,f} =\tilde{Y}_{L,f}
+2T^3_f\tilde{Y}_{\phi},
\end{eqnarray}
corresponds to the Abelian $Z^\prime$ boson. In this case the SM
Lagrangian appears to be invariant with respect to the ${\rm
\tilde{U}}(1)$ group associated with the $Z^\prime$. The first and
the second relations in Eqs. (\ref{Abelian}) mean that appropriate
generators are proportional to the unit matrix, whereas the third
relation ensures the Yukawa terms to be invariant with respect to
the ${\rm \tilde{U}}(1)$ transformations. Introducing the
$Z^\prime$ couplings to the vector and the axial-vector fermion
currents,
$v^f_{Z^\prime}\equiv(\tilde{Y}_{L,f}+\tilde{Y}_{R,f})/2$,
$a^f_{Z^\prime}\equiv(\tilde{Y}_{R,f}-\tilde{Y}_{L,f})/2$, one can
rewrite the second and the third of Eqs. (\ref{Abelian}) in the
following form:
\begin{equation}\label{Abelian1}
v^f_{Z^\prime}-a^f_{Z^\prime}
=v^{f^\star}_{Z^\prime}-a^{f^\star}_{Z^\prime},\quad
a^f_{Z^\prime} =T^3_f\tilde{Y}_{\phi}.
\end{equation}
As is seen, the couplings of the Abelian $Z^\prime$ to the
axial-vector fermion currents have a universal absolute value
proportional to the $Z^\prime$ coupling to the scalar doublets.
The solutions derived are the same as in case of the minimal SM
considered in Ref. \cite{Zpr}. Notice that both of correlations
(\ref{non-Abelian}) and (\ref{Abelian}) lead to the same
$Z^\prime$ couplings to each of the scalar doublets.

Notice, in case of the Abelian $Z^\prime$ boson the correlations
(\ref{Abelian}),(\ref{Abelian1}) can be derived on related but
formally different grounds. The point is that the
renormalizability and gauge invariance of interactions are closely
connected. Therefore, the requirement of renormalizability can be
substituted by the requirement of gauge invariance of the
effective low-energy Lagrangian.

In general, the EL respects by construction various [and, in
particular, $\tilde{U}(1)$] symmetries. But if non-renormalizable
interactions are admitted, no relations between the arbitrary
parameters can be found. If only the renormalizable interactions
are taken into account, as in Eq. (\ref{L:fermion}), some
correlations appear. In fact, to obtain formulae
(\ref{Abelian}),(\ref{Abelian1}) it is sufficient to require the
$\tilde{U}(1)$ gauge invariance of the Yukawa terms. Note also
that the correlations in Eq. (\ref{Abelian1}) are the same as in
the SM for the specific values of the hypercharges $Y_f$ and
$Y_\phi$ corresponding to the $U(1)_Y$ gauge transformations of
fermion and scalar fields. On the other hand, we did not find any
symmetry requirement describing the all possible relations
following from Eq. (\ref{non-Abelian}). Therefore, the
renormalizability requirement looks as more general one.

\section{RG correlations and the $Z^\prime$ in ${\rm E}_6$
based models} \label{sec:e6}

Over last decades the GUT's based on the ${\rm E}_6$ gauge group
\cite{hewett} are intensively studied. They predict the Abelian
$Z^\prime$ boson with the mass $m_{Z^\prime}\gg m_W$. Since the
low-energy limit of the ${\rm E}_6$ GUT's is the THDM considered,
it is of interest to check whether relations (\ref{Abelian1}) hold
for the specific values of the $Z^\prime$ couplings in these
models.

There are different schemes of the ${\rm E}_6$-symmetry breaking.
One of them is
\begin{eqnarray}\label{LR}
{\rm E}_6&\to&{\rm SO}(10)\times{\rm U}(1)_\psi,\nonumber\\
 {\rm SO}(10)&\to&{\rm SU}(3)_c\times{\rm SU}(2)_L
 \times{\rm SU}(2)_R\times\nonumber\\
 &&\times{\rm U}(1)_{B-L}.
\end{eqnarray}
This leads to the so called left-right (LR) model. Another scheme,
\begin{equation}\label{E6}
{\rm E}_6\to{\rm SO}(10)\times{\rm U}(1)_\psi\to{\rm SU}(5)\times
{\rm U}(1)_\chi\times{\rm U}(1)_\psi,
\end{equation}
predicts the Abelian $Z^\prime$, which is a linear combination of
the neutral vector bosons $\psi$ and $\chi$,
\begin{equation}\label{betadef}
Z^\prime=\chi\cos\tilde{\beta} +\psi\sin\tilde{\beta},
\end{equation}
where $\tilde{\beta}$ is the mixing angle.

In Table I (see Ref. \cite{leike}) we show the $Z^\prime$
couplings to the SM fermions in models mentioned (notice, the sign
of axial-vector couplings in Ref. \cite{leike} is opposite to the
sign of $a^f_{Z^\prime}$). At first glance, some of the couplings
in Table I are inconsistent with relations (\ref{Abelian1}). This
requires to be discussed in more detail.

First of all, let us consider the $Z^\prime$ couplings to
neutrinos. It is usually supposed in theories based on the ${\rm
E}_6$ group that the Yukawa terms responsible for generation of
the Dirac masses of neutrinos must be set to zero \cite{hewett}.
Therefore, there are no RG relations for the $Z^\prime$
interactions with the neutrino axial-vector currents, because the
terms proportional to $G_{\nu,i}$ vanish in Eq. (\ref{eq}). In
this case the couplings $a^\nu_{Z^\prime}$ given in Table I are
not restricted by relations (\ref{Abelian1}).

Now, let us discuss the $Z^\prime$ couplings to charged leptons
and quarks. The values of the couplings satisfy relations
(\ref{Abelian1}) in case of the LR model. As for models described
by the ${\rm E}_6$ breaking scheme (\ref{E6}), two possibilities
of choosing $\tilde{\beta}$ are of interest. First is if the
$\psi$ boson is much heavier than the $\chi$ field. In general,
this is a natural condition, since the fields $\psi$ and $\chi$
arise at different energy scales. As a consequence, the field
$\psi$ is decoupled, and the mixing angle $\tilde{\beta}$ is small
($\tilde{\beta}\ll 1$). In this case RG relations (\ref{Abelian1})
hold in lower order in $\tilde{\beta}$ for the $Z^\prime$
couplings to quarks and charged leptons.

The second possibility is if the masses of $\chi$ and $\psi$ are
of the same order. This means the tuning of the vacuum expectation
values generating the vector boson masses. This case cannot be
treated straightforwardly on the basis of relations
(\ref{Abelian1}) since the mixed states of the $Z^\prime$ bosons
have to be included into consideration explicitly. Although our
approach is applicable in this case, it requires additional
investigation. Moreover, the $Z^\prime$ mixed states cause some
different exchange amplitudes, which have to be incorporated into
low-energy observables. In what follows, we will not discuss the
case of two $Z^\prime$ bosons having masses of the same order.

\section{Observables}\label{sec:observ}

Now, let us introduce the observables convenient for detection of
the $Z^\prime$ in electron-positron annihilation into fermion
pairs $e^+e^-\to V^\ast\to\bar{f}f$ ($f\not=e,t$). The
center-of-mass energy is taken in the range $\sqrt{s}\ge 500$ GeV.
Consider the case of non-polarized initial and final fermions.
Since the $t$ quark is not considered, other fermions at these
energies can be treated as massless particles, $m_f\sim 0$. In
this approximation the left-handed and the right-handed fermions
can be substituted by the helicity states, which will be marked as
$\lambda$ and $\xi$ for the incoming electron and the outgoing
fermion, respectively ($\lambda,\xi=L,R$).

Let ${\cal A}_V$ be the Born amplitude of the process $e^+e^-\to
V^\ast\to\bar{f}f$ ($f\not=e,t$) with the virtual $V$-boson state
in the $s$ channel ($V=A,Z,Z^\prime$). The $Z^\prime$ boson
existence leads to the deviation of order $\sim m^{-2}_{Z^\prime}$
of the cross section from its SM value. In general, the tree-level
deviations originate from two types of contributions. The first is
caused by the $Z$-$Z^\prime$ mixing. Using the results of Sec.
\ref{sec:RGrelations} the mixing angle $\theta_0$ [see Eq.
(\ref{ZZpmixing})] can be calculated as follows,
\begin{eqnarray}
\theta_0&\simeq& \frac{\tilde{g} m^2_W
\tilde{Y}_\phi}{g\cos{\theta_W} m^2_{Z^\prime}}.
\end{eqnarray}
Because of the mixing there are corrections of order $\theta_0\sim
m^{-2}_{Z^\prime}$ to the vertex describing interaction of $Z$
boson and fermions. Hence, the amplitude ${\cal A}_Z(\theta_0)$
deviates from its SM value ${\cal A}_Z(\theta_0=0)$. The second
type describes the interference between the SM amplitude, ${\cal
A}_{\rm SM}$, and the $Z^\prime$ exchange amplitude, ${\cal
A}_{Z^\prime}$. Thus, for the process $e^+e^-\to\bar{f}f$ the
deviation of the cross section is
\begin{equation}\label{crosssection}
\Delta\frac{d\sigma_f}{d\Omega}
 =\frac{d\sigma_f}{d\Omega}-\frac{d\sigma_{f,\rm SM}}{d\Omega}
 =\frac{\mbox{Re}\left[{\cal A}^\ast_{\rm SM}\Delta{\cal A}\right]}
 {32\pi s} +O\left(\frac{s^2}{m^4_{Z^\prime}}\right),
\end{equation}
where
\begin{equation}\label{amplitude}
{\cal A}_{\rm SM}={\cal A}_A+\left.{\cal A}_Z\right|_{\theta_0
=0},~~ \Delta{\cal A} ={\cal A}_{Z^\prime} +\left(\frac{d{\cal
A}_Z}{d\theta_0}\right)_{\theta_0 =0} \theta_0.
\end{equation}

The quantity $\Delta d\sigma/d\Omega$ can be calculated in the
form
\begin{equation}\label{deviation}
 \Delta\frac{d\sigma_f}{d\Omega}=\sum\limits_{\lambda,\xi=L,R}
 \left[{\cal I}^{ef}_{\lambda\xi}(s)
 +{\cal M}^{ef}_{\lambda\xi}(s)\right]
 {\left(z +P_\lambda P_\xi\right)}^2,
\end{equation}
where $P_L=-1$, $P_R=1$, $z\equiv\cos\theta$ ($\theta$ is the
angle between the incoming electron and the outgoing fermion),
${\cal I}^{ef}_{\lambda\xi}$ denotes the $Z$-$Z^\prime$
interference term, and ${\cal M}^{ef}_{\lambda\xi}$ accounts of
the contributions from the $Z$-$Z^\prime$ mixing:
\begin{eqnarray}\label{deviation1}
 {\cal I}^{ef}_{\lambda\xi}&=&
 \frac{\alpha_{\rm em}\tilde{g}^2 T^3_f N_f}
  {4\pi m^2_{Z^\prime}}
 \tilde{Y}_{\lambda,e}\tilde{Y}_{\xi,f}
 \left[|Q_f|
 \right.\nonumber\\&&
   +\left.\chi(s)\left(P_\lambda -\varepsilon\right)
   \left(P_\xi -1 +|Q_f| -|Q_f|\varepsilon\right)
 \right],
 \nonumber\\
 {\cal M}^{ef}_{\lambda\xi}&=&
 \frac{\alpha_{\rm em}g\tilde{g} T^3_f N_f \theta_0}
  {4\pi\cos{\theta_W}(s -m^2_Z)}
 \left[\tilde{Y}_{\xi,f}
   \left(\delta_{\lambda,L}-2\sin^2{\theta_W}\right)
 \right.\nonumber\\&&+\left.
   2T^3_f\tilde{Y}_{\lambda,e}
    \left(2|Q_f|{\sin}^2{\theta_W}-\delta_{\xi,L}\right)
 \right]
 \left[|Q_f|
 \right.\nonumber\\&&
   +\left.\chi(s)\left(P_\lambda -\varepsilon\right)
   \left(P_\xi -1 +|Q_f| -|Q_f|\varepsilon\right)
   \right],
\end{eqnarray}
where $\alpha_{\rm em}$ is the fine structure constant, $N_f=3$
for quarks and $N_f=1$ for leptons, $\varepsilon\equiv
1-4\sin^2\theta_W \sim 0.08$,
$\chi^{-1}(s)=16\sin^2\theta_W\cos^2\theta_W(1-m^2_Z/s)$, and
$\delta_{\lambda,\xi}$ is the Kronecker delta. The leading
contribution comes from the $Z$-$Z^\prime$ interference term
${\cal I}^{ef}_{\lambda\xi}$, whereas the $Z$-$Z^\prime$ mixing
terms are suppressed by the additional factor $m^2_Z/s$. At
energies $\sqrt{s}\ge 500$ GeV ${\cal M}^{ef}_{\lambda\xi}\ll{\cal
I}^{ef}_{\lambda\xi}$.

To take into consideration the correlations (\ref{non-Abelian}) or
(\ref{Abelian}) let us introduce the function $\sigma_f(z)$
defined as the difference of cross sections integrated in a
suitable range of $\cos\theta$ \cite{obs}:
\begin{equation}\label{sigmaz}
 \sigma_f(z)
 \equiv\int\nolimits_z^1\frac{d\sigma_f}{dz}dz
 -\int\nolimits_{-1}^z\frac{d\sigma_f}{dz}dz.
\end{equation}
The conventionally used observables -- the total cross section
$\sigma_{f,T}$ and the forward-backward asymmetry $A_{f,FB}$ --
can be obtained by a special choice of $z$ [$\sigma_{f,T}
=\sigma_f(-1)$, $A_{f,FB}=\sigma_f(0)/\sigma_{f,T}$]. One can
express $\sigma_f(z)$ in terms of $\sigma_{f,T}$ and $A_{f,FB}$:
\begin{equation}\label{traditional}
\sigma_f(z) =\sigma_{f,T}\left[ A_{f,FB}\left(1-z^2\right)
-\frac{1}{4}z\left(3+z^2\right)\right].
\end{equation}

Then, the deviation $\Delta\sigma_f(z)\equiv\sigma_f(z)
-\sigma_{f,\rm SM}(z)$ can be written in the form:
\begin{eqnarray}\label{observable}
\Delta\sigma_f(z)&=&
 4\pi\sum\limits_{\lambda,\xi}
 \left[{\cal I}^{ef}_{\lambda\xi}(s)
 +{\cal M}^{ef}_{\lambda\xi}(s)\right]
 \nonumber\\&&\times
 \left(P_\lambda P_\xi -z
  -z^2 P_\lambda P_\xi -\frac{z^3}{3}\right).
\end{eqnarray}
Let us compare the observable $\Delta\sigma_f(z)$ with the
differential cross section (\ref{deviation}). As is seen, the
polynomial in the polar angle $z$ in Eq. (\ref{deviation}) is
replaced by the function of the boundary angle $z$ in Eq.
(\ref{observable}). The overall factor $4\pi$ appears due to the
angular integration.

In what follows, we consider the observable (\ref{observable})
taking into account correlations (\ref{non-Abelian}) and
(\ref{Abelian}).

\subsection{Chiral $Z^\prime$}

The case of the chiral $Z^\prime$ corresponds to correlations
(\ref{non-Abelian}). Because of absence of the $Z^\prime$
couplings to right-handed fermions the leading contribution to
$\Delta\sigma_f(z)$ is proportional to the same polynomial in $z$
for any outgoing fermion $f$:
\begin{eqnarray}\label{nonAbeliansigma}
 \Delta\sigma_f(z)
 &\simeq& 4\pi{\cal I}^{ef}_{LL}(s)
 \left(1 -z -z^2 -\frac{z^3}{3}\right)\nonumber\\
 &=& \frac{\alpha_{\rm em}\tilde{g}^2 T^3_f N_f}{m^2_{Z^\prime}}
 \tilde{Y}_{L,e}\tilde{Y}_{L,f}
 \left(1 -z -z^2 -\frac{z^3}{3}\right)
 \nonumber\\&&
 \times\left\{\left[|Q_f|+2\chi(s) -|Q_f|\chi(s)\right]
  +O\left(\varepsilon\right)\right\}.
\end{eqnarray}
Therefore, the differential cross section is completely determined
by the total one:
\begin{eqnarray}
 \Delta\sigma_f(z)&=&\Delta\sigma_{f,T}
  \Big[\frac{3}{4}\left(1 -z -z^2 -\frac{z^3}{3}\right)
 \nonumber\\&&
  +O\left(\varepsilon,m^2_Z s^{-1}\right)\Big].
\end{eqnarray}

Comparing the observables for fermions of the same ${\rm SU}(2)_L$
isodoublet, $\{ f_u, f_d\}$, it is possible to derive the
correlation:
\begin{eqnarray}
 \Delta\sigma_{f_u}(z)&=& \Delta\sigma_{f_d}(z)\left[
 \frac{|Q_{f_u}|+1}{|Q_{f_d}|+1}
 +O\left(\varepsilon,m^2_Z s^{-1}\right)\right].
\end{eqnarray}
Hence, the ratio $\Delta\sigma_{f_u}(z)/ \Delta\sigma_{f_d}(z)$ is
independent of $z$. It equals to 5/4 for quarks and 1/2 for
leptons in lower order in $\varepsilon$, $m^2_Z s^{-1}$. So, the
values of the observables in the $\Delta\sigma_{f_u}(z)$ --
$\Delta\sigma_{f_d}(z)$ plane are at the same curve (straight line
in the approximation used) for any $z$ specified.

It also follows from Eq. (\ref{nonAbeliansigma}) that there is a
value $z=z^\prime$ when $\Delta\sigma(z^\prime)=0$. As one can
check, $z^\prime =2^{2/3}-1$. Notice, the observable
$\Delta\sigma(z^\prime)$ is just the variable $\Delta\sigma_-$
proposed in Ref. \cite{PPO}. This quantity is completely
insensitive to the chiral $Z^\prime$ signals. On the other hand,
the deviation of the total cross section, $\Delta\sigma_T$, is
more sensitive to signals of the chiral $Z^\prime$, since the
maximum of the polynomial $1 -z -z^2 -z^3/3$ is at $z=-1$.

\subsection{Abelian $Z^\prime$}

The Abelian $Z^\prime$ beyond the minimal SM was considered
recently in Ref. \cite{obs}, where sign-definite observables
convenient for detection of the Abelian $Z^\prime$ have been
introduced. RG correlations (\ref{Abelian}) in Sec.
\ref{sec:RGrelations} coincide with that of Ref. \cite{obs}.
Therefore, the observables for Abelian $Z^\prime$ beyond the THDM
are to be the same as in case of the minimal SM.

\begin{figure}
\begin{center}
 \epsfxsize=0.35\textwidth
 \epsfbox[0 0 600 600]{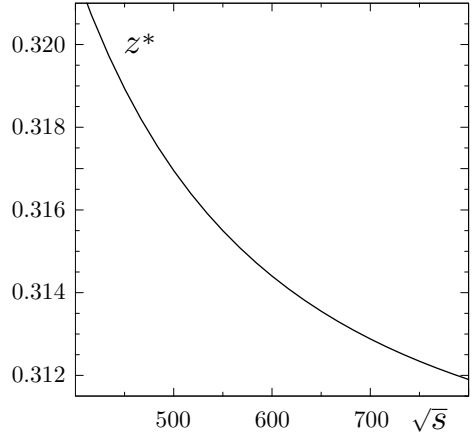}
 \caption{$z^\ast$ as the function of $\sqrt{s}$(GeV).}
 \label{fig:3}
 \end{center}
\end{figure}

In case of the chiral $Z^\prime$ the RG correlations
(\ref{non-Abelian}) suppress amplitudes corresponding to the
processes with right-handed fermions. This is not the case for the
Abelian $Z^\prime$. However, one can switch off some contributions
to observable (\ref{observable}) by specifying the kinematic
parameter $z$. In what follows, it will be convenient to use the
$Z^\prime$ couplings to vector and axial-vector fermion currents
[$v^f_{Z^\prime}\equiv(\tilde{Y}_{L,f}+\tilde{Y}_{R,f})/2$,
$a^f_{Z^\prime}\equiv(\tilde{Y}_{R,f}-\tilde{Y}_{L,f})/2$].
Because of correlations (\ref{Abelian1}) the absolute value of the
axial-vector couplings is universal for the all types of SM
fermions, $a_{Z^\prime}\sim\tilde{Y}_{\phi}$. So, the observable
$\Delta\sigma_f(z)$ has the form
\begin{eqnarray}
\Delta\sigma_f(z)&=& \frac{\alpha_{\rm
em}\tilde{g}^2}{m^2_{Z^\prime}} \left[{\cal
F}^f_0(z,s)a^2_{Z^\prime} + {\cal F}^f_1(z,s)v^e_{Z^\prime}
v^f_{Z^\prime}
 \right.\nonumber\\&&\left.
+{\cal F}^f_2(z,s)a_{Z^\prime} v^f_{Z^\prime} +{\cal
F}^f_3(z,s)v^e_{Z^\prime} a_{Z^\prime}\right].
\end{eqnarray}
As it was argued in Ref. \cite{obs}, one is able to choose the
value of $z=z^\ast$, which switches off the leading contributions
to the leptonic factors ${\cal F}^l_1$, ${\cal F}^l_2$, and the
factor ${\cal F}^f_3$. The appropriate value of $z^\ast$ can be
found from the equation
\begin{eqnarray}\label{zAbelian}
\chi(s)\left(1 -z^{\ast 2}\right)
 -\left(z^\ast +\frac{z^{\ast 3}}{3}\right)
 \left[1 +\chi(s)\varepsilon^2\right]&=& 0.
\end{eqnarray}
The solution $z^\ast(s)$ is shown in Fig. \ref{fig:3}. This
switches off the factor at $v^e_{Z^\prime} v^l_{Z^\prime}$. As is
seen, $z^\ast$ decreases from 0.317 at $\sqrt{s}=500$ GeV to 0.313
at $\sqrt{s}=700$ GeV. In what follows the value of $\sqrt{s}$ is
taken to be 500 GeV, because $z^\ast$ and $\Delta\sigma(z)$ depend
on the center-of-mass energy through the small quantity $m^2_Z/s$
(such contributions are of order 3\%).

With the above discussed choice of $z^\ast$ made, one can
introduce the sign definite observable $\Delta\sigma_l(z^\ast)$:
\begin{eqnarray}\label{obs:9}
 \Delta\sigma_l(z^\ast)&=&
 \frac{\alpha_{\rm em}\tilde{g}^2}{m^2_{Z^\prime}}
 {\cal F}^l_0(z^\ast,s)a^2_{Z^\prime} \nonumber\\
 &=&-0.10\frac{\alpha_{\rm em}\tilde{g}^2\tilde{Y}^2_\phi}
 {m^2_{Z^\prime}}\left[1 +O\left(0.04\right)\right]<0.
\end{eqnarray}
Notice, the value of $\Delta\sigma_l(z^\ast)$ is universal for the
all types of SM charged leptons. There are also sign definite
observables for the quarks of the same generation:
\begin{equation}\label{obs:10a}
\Delta\sigma_q(z^\ast) \equiv\Delta\sigma_{q_u} +0.5
\Delta\sigma_{q_d} \simeq 2.45\Delta\sigma_l\left(z^\ast\right)<0.
\end{equation}
Hence one can conclude that the values of
$\Delta\sigma_{q_u}(z^\ast)$ and $\Delta\sigma_{q_d}(z^\ast)$ in
the $\Delta\sigma_{q_u}(z^\ast)$ -- $\Delta\sigma_{q_d}(z^\ast)$
plane have to be at the line crossing the axes at the points
$\Delta\sigma_{q_u}(z^\ast) =2.45\Delta\sigma_l(z^\ast)$ and
$\Delta\sigma_{q_d}(z^\ast) =4.9\Delta\sigma_l(z^\ast)$,
respectively.

\begin{figure}
\begin{center}
 \epsfxsize=0.35\textwidth
 \epsfbox[0 0 600 600]{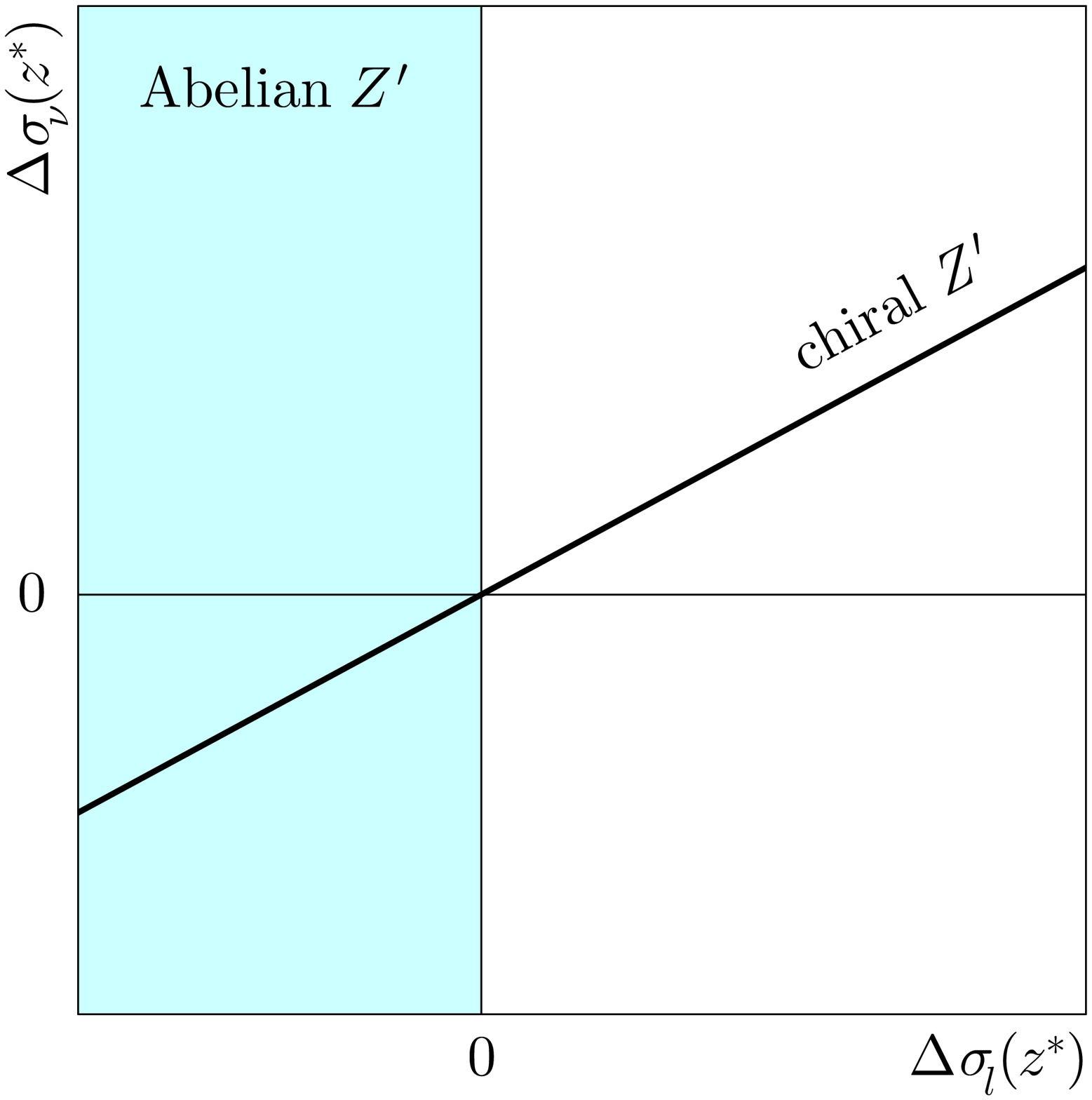}
 \caption{Signals of the Abelian and the chiral $Z^\prime$ in the
  plane of observables $\Delta\sigma_{q_l}(z^\ast)$ and
  $\Delta\sigma_{q_{\nu_l}}(z^\ast)$ for leptons of the same
  generation. The shaded area represents the signal of the Abelian
  $Z^\prime$ for all possible values of the axial-vector couplings
  $a^f_{Z^\prime}$.}
 \label{fig:4}
 \end{center}
\end{figure}

\begin{figure}
\begin{center}
 \epsfxsize=0.35\textwidth
 \epsfbox[0 0 600 600]{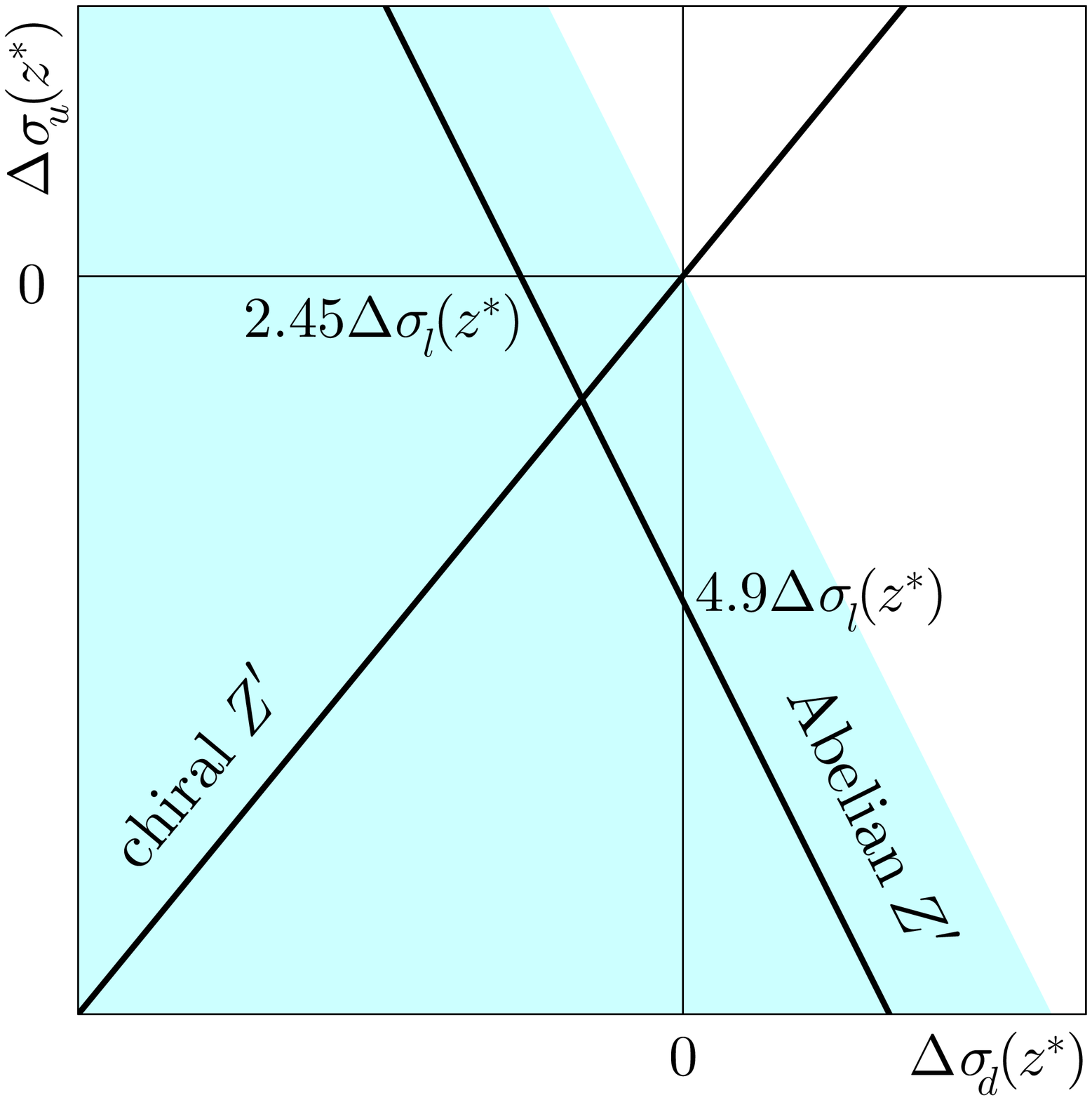}
 \caption{Signals of the Abelian and the chiral $Z^\prime$ in the
  plane of observables $\Delta\sigma_{q_d}(z^\ast)$ and
  $\Delta\sigma_{q_u}(z^\ast)$ for quarks of the same generation.
  The shaded area represents the signal of the Abelian $Z^\prime$
  for all possible values of the axial-vector couplings
  $a^f_{Z^\prime}$.}
 \label{fig:5}
 \end{center}
\end{figure}

Signals of the Abelian and the chiral $Z^\prime$ are compared in
Figs. \ref{fig:4}-\ref{fig:5}. Suppose for a moment that
experiments give the non-zero values of leptonic observables
$\Delta\sigma_l(z^\ast)$ ($l=\mu,\tau$). If they correspond to the
Abelian $Z^\prime$, either of the observables has to be the same
negative number. Let one also know the values of the neutrino
observables $\Delta\sigma_\nu(z^\ast)$ ($\nu=\nu_\mu,\nu_\tau$).
In case of the chiral $Z^\prime$ the corresponding point in Fig.
\ref{fig:4} has to be at the straight line shown (with the
accuracy of the approximation). For the Abelian $Z^\prime$ the
shaded region as a whole is available. Now, let us consider
observables for the quarks of the same generation (see Fig.
\ref{fig:5}). If the value of the leptonic observable
$\Delta\sigma_l(z^\ast)$ is measured, one has to expect that the
experimental points will be located at one of two possible curves
corresponding either to the chiral or to the Abelian $Z^\prime$.
The shaded range represents signals of the Abelian $Z^\prime$ for
the all possible values of the leptonic observable. So, by
measuring the observables $\Delta\sigma_f(z^\ast)$ for fermions of
the same ${\rm SU}(2)_L$ isodoublet, one is able to distinguish
the Abelian and the chiral $Z^\prime$ couplings.

\section{Discussion}\label{sec:discussion}

In the present paper the method of RG relations \cite{Zpr,hhiggs},
developed originally for the minimal SM, is extended to searching
for signals of the heavy $Z^\prime$ gauge boson beyond the THDM.
General conditions when our consideration is applicable are the
following. 1) The mechanism generating the heavy particle masses
is not specified, and the $Z^\prime$ mass is treated as an
arbitrary parameter. 2) The light particle masses are generated in
a standard way via the non-zero vacuum values of the scalar fields
of the low-energy basis theory. Interactions of light particles
with heavy scalar fields, which are responsible for
$m_{Z^\prime}$, are excluded at tree level. The radiation
corrections to the masses due to heavy particle loops are
suppressed by factors $\sim O(m_{\rm light}/m_{Z^\prime})$, and
therefore not taken into account. This kind of the mass hierarchy
corresponds to the case when the basis theory is a subgroup of the
underlying high energy model remaining unknown.

As our consideration shown, only two types of the $Z^\prime$
couplings to light particles are consistent with the
renormalizability. The first type corresponds to the Abelian
couplings respecting the ${\rm \tilde{U}}(1)$ symmetry of the
effective Lagrangian (\ref{L:fermion}). In this case, the RG
correlations fix the gauge symmetry of the Yukawa terms, which
relates the fermion and the scalar hypercharges. As a consequence,
the $Z^\prime$ couplings to the axial-vector fermion currents are
completely determined by the scalar field hypercharge and the
fermion isospin. The second set of solutions -- chiral $Z^\prime$
-- describes interactions with the SM particles similar to the
third component of the ${\rm SU}(2)_L$ gauge field. The
characteristic feature of the latter couplings is the zero traces
of generators associated with the $Z^\prime$. Notice that the
$Z^\prime$ interactions of the chiral type result in the effective
four-fermion couplings $(\bar{f}_{1L}\gamma^\mu\sigma^a
f_{1L})(\bar{f}_{2L}\gamma^\mu\sigma^a f_{2L})$ described by the
operators ${\cal O}^{(3)}_{ll}$, ${\cal O}^{(3)}_{lq}$, and ${\cal
O}^{(1,3)}_{qq}$ according to the classification in Refs.
\cite{operators}. Since each type of the $Z^\prime$ interactions
corresponds to one of mentioned operators, there is a possibility
to select interactions by constructing the proper observables. As
was shown, the observables proposed in Ref. \cite{obs} can be
chosen in searching for the Abelian $Z^\prime$ boson. Thus, the
bounds on the $Z^\prime$ couplings calculated therein are also
applicable in case of the THDM.

The above note is important for the model independent search for
$Z^\prime$ virtual states at LEP2 and future colliders LHC and
NLC. In the analysis of experimental data no discriminations
between these two cases have been discussed in literature (see,
for instance, recent survey \cite{leike} or report
\cite{riemann}). This difference should be important for the
model-dependent $Z^\prime$ search when different scenarios of
symmetry breaking are discussed.

We believe that the derived RG relations to be useful in improving
of experimental bounds on either the parameters of the $Z^\prime$
interaction with fermions and on the relations between the cross
sections of various four-fermion scattering processes.

\section*{Acknoledgments}

The authors thank S. V. Peletminski and N. F. Shul'ga for
discussions.

\appendix
\section{Feynman rules}\label{sec:vertices}

In what follows we use the notation
$\omega_{L,R}=(1\mp\gamma^5)/2$, and all the momenta in the
vertices are understood to be incoming. The Feynman rules for
vertices of Figs. \ref{fig:1}, \ref{fig:2} are listed below:
\begin{enumerate}
\item Fermion-vector vertices
\begin{eqnarray}
\bar{f}f A_\mu :&\quad& g\sin{\theta_W}Q_f\gamma^\mu;
 \nonumber\\
\bar{f}f Z_\mu :&\quad& \frac{g}{\cos{\theta_W}}\gamma^\mu
  \left(T^3_f{\omega_L} -{Q_f}\sin^2{\theta_W}\right)
 \nonumber\\
 && +O(\theta_0);
 \nonumber\\
\bar{f}f Z^\prime_\mu :&\quad& \frac{\tilde{g}}{2}\gamma^\mu
 \left(\omega_L\tilde{Y}_{L,f} +\omega_R\tilde{Y}_{R,f}\right)
 +O(\theta_0);
 \nonumber\\
\bar{f}_d f_u W^-_\mu :&\quad& \frac{g}{\sqrt{2}}\gamma^\mu
 \omega_L;
 \nonumber\\
\bar{f}_u f_d W^+_\mu :&\quad& \frac{g}{\sqrt{2}}\gamma^\mu
 \omega_L;
 \nonumber
\end{eqnarray}
\item Fermion-scalar vertices
\begin{eqnarray}
\bar{f}f H :&\quad&
 -\left(G_{f,1}\cos\alpha +G_{f,2}\sin\alpha\right);
 \nonumber\\
\bar{f}f h :&\quad&
 \left(G_{f,1}\sin\alpha -G_{f,2}\cos\alpha\right);
 \nonumber\\
\bar{f}f A_0 :&\quad&
 2i T^3_f \left(\omega_L -\omega_R\right)
 \nonumber\\
 &&\times \left(G_{f,1}\sin\beta -G_{f,2}\cos\beta\right);
 \nonumber\\
\bar{f}f \chi_3 :&\quad&
 -2i T^3_f \left(\omega_L -\omega_R\right)
 \nonumber\\
 &&\times \left(G_{f,1}\cos\beta +G_{f,2}\sin\beta\right);
 \nonumber\\
\bar{f}_d f_u H^- :&\quad& \sqrt{2}\left[
 \omega_L\left(G_{f_d,1}\sin\beta -G_{f_d,2}\cos\beta\right)
 \right.\nonumber\\&&\left.
 +\omega_R\left(-G_{f_u,1}\sin\beta +G_{f_u,2}\cos\beta\right)
 \right];
 \nonumber\\
\bar{f}_u f_d H^+ :&\quad& \sqrt{2}\left[
 \omega_R\left(G_{f_d,1}\sin\beta -G_{f_d,2}\cos\beta\right)
 \right.\nonumber\\&&\left.
 +\omega_L\left(-G_{f_u,1}\sin\beta +G_{f_u,2}\cos\beta\right)
 \right];
 \nonumber\\
\bar{f}_d f_u \chi^- :&\quad& \sqrt{2}\left[
 -\omega_L\left(G_{f_d,1}\cos\beta +G_{f_d,2}\sin\beta\right)
 \right.\nonumber\\&&\left.
 +\omega_R\left(G_{f_u,1}\cos\beta +G_{f_u,2}\sin\beta\right)
 \right];
 \nonumber\\
\bar{f}_u f_d \chi^+ :&\quad& \sqrt{2}\left[
 -\omega_R\left(G_{f_d,1}\cos\beta +G_{f_d,2}\sin\beta\right)
 \right.\nonumber\\&&\left.
 +\omega_L\left(G_{f_u,1}\cos\beta +G_{f_u,2}\sin\beta\right)
 \right];
 \nonumber
\end{eqnarray}
\item $Z^\prime$ scalar vertices
\begin{eqnarray}
Z^\prime_\mu H^+ H^- :&\quad& \frac{\tilde{g}}{2}
 \left(p_{H^+} -p_{H^-}\right)_\mu
 \left(\tilde{Y}_{\phi_1,1}\sin^2{\beta}
 \right.\nonumber\\&&\left.
  +\tilde{Y}_{\phi_2,1}\cos^2{\beta}\right)
 +O(\theta_0);
 \nonumber\\
Z^\prime_\mu H^+ \chi^- :&\quad& \frac{\tilde{g}\sin{2\beta}}{4}
 \left(p_{\chi^-} -p_{H^+}\right)_\mu
 \nonumber\\
 &&\times
 \left(\tilde{Y}_{\phi_1,1} -\tilde{Y}_{\phi_2,1}\right)
 +O(\theta_0);
 \nonumber\\
Z^\prime_\mu H^- \chi^+ :&\quad& \frac{\tilde{g}\sin{2\beta}}{4}
 \left(p_{H^-} -p_{\chi^+}\right)_\mu
 \nonumber\\
 &&\times
 \left(\tilde{Y}_{\phi_1,1} -\tilde{Y}_{\phi_2,1}\right)
 +O(\theta_0);
 \nonumber\\
Z^\prime_\mu \chi^+ \chi^- :&\quad& \frac{\tilde{g}}{2}
 \left(p_{\chi^+} -p_{\chi^-}\right)_\mu
 \left(\tilde{Y}_{\phi_1,1}\cos^2{\beta}
 \right.\nonumber\\&&\left.
  +\tilde{Y}_{\phi_2,1}\sin^2{\beta}\right)
 +O(\theta_0);
 \nonumber\\
Z^\prime_\mu H A_0 :&\quad& \frac{i\tilde{g}}{2}
 \left(p_{A_0}-p_H\right)_\mu
 \tilde{Y}_{\phi,2}\sin\left(\alpha -\beta\right)
 \nonumber\\&&
 +O(\theta_0);
 \nonumber\\
Z^\prime_\mu H \chi_3 :&\quad& \frac{i\tilde{g}}{2}
 \left(p_{\chi_3} -p_H\right)_\mu
 \tilde{Y}_{\phi,2}\cos\left(\alpha -\beta\right)
 \nonumber\\&&
 +O(\theta_0);
 \nonumber\\
Z^\prime_\mu h A_0 :&\quad& \frac{i\tilde{g}}{2}
 \left(p_{A_0} -p_h\right)_\mu
 \tilde{Y}_{\phi,2}\cos\left(\alpha -\beta\right)
 \nonumber\\&&
 +O(\theta_0);
 \nonumber\\
Z^\prime_\mu h \chi_3 :&\quad& \frac{i\tilde{g}}{2}
 \left(p_h -p_{\chi_3}\right)_\mu
 \tilde{Y}_{\phi,2}\sin\left(\alpha -\beta\right)
 \nonumber\\&&
 +O(\theta_0).
 \nonumber
\end{eqnarray}
\end{enumerate}

\newpage
\begin{table}
\begin{center}
\label{zpcoup} \caption{The $Z^\prime$ couplings to the SM
fermions in the ${\rm E}_6$ and LR models.}
\begin{tabular}{|l|lcc|lcc|}
$f$&${\rm E}_6$:& $a^f_{Z^\prime}$ & $v^f_{Z^\prime}$
    & LR:    & $a^f_{Z^\prime}$ & $v^f_{Z^\prime}$
\\ \hline
&&&&&&\\ $\nu$& &
 $-3\frac{\cos\tilde{\beta}}{\sqrt{40}}
  -\frac{\sin\tilde{\beta}}{\sqrt{24}}$ &
 $3\frac{\cos\tilde{\beta}}{\sqrt{40}}
  +\frac{\sin\tilde{\beta}}{\sqrt{24}}$ & &
 $-\frac{1}{2\alpha}$ & $\frac{1}{2\alpha}$ \\
&&&&&&\\
 $e$ & &
 $-\frac{\cos\tilde{\beta}}{\sqrt{10}}
  -\frac{\sin\tilde{\beta}}{\sqrt{6}}$ &
 $2\frac{\cos\tilde{\beta}}{\sqrt{10}}$ & &
 $-\frac{\alpha}{2}$ & $\frac{1}{\alpha}-\frac{\alpha}{2}$ \\
&&&&&&\\
 $u$ & &
 $\frac{\cos\tilde{\beta}}{\sqrt{10}}
 -\frac{\sin\tilde{\beta}}{\sqrt{6}}$ & 0 & &
 $\frac{\alpha}{2}$&$-\frac{1}{3\alpha}+\frac{\alpha}{2}$ \\
&&&&&&\\
 $d$ & &
 $-\frac{\cos\tilde{\beta}}{\sqrt{10}}
  -\frac{\sin\tilde{\beta}}{\sqrt{6}}$ &
 $-2\frac{\cos\tilde{\beta}}{\sqrt{10}}$ & &
 $-\frac{\alpha}{2}$ & $-\frac{1}{3\alpha}-\frac{\alpha}{2}$
\\&&&&&&
\end{tabular}
\end{center}
\end{table}


\begin{thebibliography}{99}
\bibitem{leike} A. Leike, Phys. Rep. {\bf 317}, 143 (1999).
\bibitem{cvetic}M. Cveti\v{c} and B. W. Lynn,
 Phys. Rev. D {\bf 35}, 51 (1987).
\bibitem{riemann} S. Riemann, in
 {\it Beyond the Standard Model V, Balholm, Norway,
 April-May 1997} (AIP Conference Proceedings 415), p. 387.
\bibitem{sirlin} G. Degrassi and A. Sirlin,
 Phys. Rev. D {\bf 40}, 3066 (1989).
\bibitem{decoupling} T. Appelquist and J. Carazzone,
 Phys. Rev. D {\bf 11}, 2856 (1975); \\
 J. C. Collins, F. Wilczek, and A. Zee,
 {\it ibid.} {\bf 18}, 242 (1978).
\bibitem{bando} M. Bando, T. Kugo, N. Maekawa, and H. Nakano,
 Progress of Theor. Phys. {\bf 90}, 405 (1993);
 Phys. Lett. B {\bf 301}, 83 (1993).
\bibitem{wudka} J. Wudka,
 Int. J. of Modern Phys. A {\bf 9}, 2301 (1994).
\bibitem{Zpr} A. V. Gulov and V. V. Skalozub, hep-ph/9812485.
\bibitem{obs} A. V. Gulov and V. V. Skalozub,
 Phys. Rev. D {\bf 61}, 055007 (2000).
\bibitem{hhguide} J. Gunion, H. Haber, G. Kane, and S. Dawson,
 {\it The Higgs Hunter's Guide} (Addison-Wesley, Reading, MA, 1990)
\bibitem{santos} R. Santos and A. Barroso,
 Phys. Rev. D {\bf 56}, 5366 (1997).
\bibitem{caso} C. Caso {\it et al.},
 Eur. Phys. J. C {\bf 3}, 1 (1998).
\bibitem{FCNC} S. Glashow and S. Weinberg,
 Phys. Rev. D {\bf 15}, 1958 (1977).
\bibitem{2sc} A. V. Gulov and V. V. Skalozub,
 Yad. Fiz. {\bf 63}, No.1 (2000).
\bibitem{hewett} J. Hewett and T. Rizzo,
 Phys. Rep. {\bf 183} (1989), 193.
\bibitem{PPO} P. Osland and A. Pankov,
 Phys. Lett. B {\bf 406}, 328 (1997);
 A. Pankov and N. Paver, {\it ibid.} {\bf 432}, 159 (1998);
 A. Babich, A. Pankov, and N. Paver,
 {\it ibid.} {\bf 426}, 375 (1998).
\bibitem{hhiggs} A. V. Gulov and V. V. Skalozub,
 Yad. Fiz. {\bf 62}, 341 (1999)
 [Phys. At. Nucl. {\bf 62}, 306 (1999)].
\bibitem{operators} W. Buchm\"uller and D. Wyler,
 Nucl. Phys. B {\bf 268}, 621 (1986);
 C. Arzt, M. Einhorn, and J. Wudka,
 {\it ibid.} {\bf 433}, 41 (1995).
\end{thebibliography}
\end{document}